\documentclass[12pt]{iopart}

\usepackage{graphicx}
\usepackage{iopams}

\begin{document}


\paper[A high quality, efficiently coupled microwave cavity for trapping cold molecules]{A high quality, efficiently coupled microwave cavity for trapping cold molecules}

\author{D P Dunseith\footnote{These two authors contributed equally to this work.}, S Truppe\footnotemark[1], R J Hendricks, B E Sauer, E A Hinds and M R Tarbutt}

\address{Centre for Cold Matter, Blackett Laboratory, Imperial College London, Prince Consort Road, London SW72BW, UK}

\ead{m.tarbutt@imperial.ac.uk}

\begin{abstract}
We characterize a Fabry-P\'erot microwave cavity designed for trapping atoms and molecules at the antinode of a microwave field. The cavity is fed from a waveguide through a small coupling hole. Focussing on the compact resonant modes of the cavity, we measure how the electric field profile, the cavity quality factor, and the coupling efficiency, depend on the radius of the coupling hole. We measure how the quality factor depends on the temperature of the mirrors in the range from 77 to 293\,K. The presence of the coupling hole slightly changes the profile of the mode, leading to increased diffraction losses around the edges of the mirrors and a small reduction in quality factor. We find the hole size that maximizes the intra-cavity electric field. We develop an analytical theory of the aperture-coupled cavity that agrees well with our measurements, with small deviations due to enhanced diffraction losses. We find excellent agreement between our measurements and finite-difference time-domain simulations of the cavity.

\end{abstract}

\maketitle


\section{Introduction}

There is currently great interest in cooling a wide variety of molecules to low temperatures and controlling both the internal states and the external motion of these cold molecules \cite{Carr:2009}. This interest is motivated by a diverse range of applications in physics and chemistry. These include precise molecular spectroscopy to test new theories of physics \cite{Hudson2011, ACME2014, Truppe2013, DeMille2008, Darquie2010}, investigating the physics of strongly-interacting many-body quantum systems \cite{Goral2002, Micheli2006}, studying and controlling collisions and reactions at low temperatures \cite{Krems2008}, and quantum information processing \cite{DeMille2002, Andre2006}.

Molecules in weak-field seeking states can be trapped using static electric or magnetic fields \cite{Bethlem2000, Weinstein1998, Sawyer2007} but these traps cannot hold ground-state molecules which are always strong-field seeking. It is important to trap ground-state molecules, particularly for sympathetic cooling and evaporative cooling schemes that rely on collisions to cool the molecules. Unless they are in the ground state, inelastic collisions will tend to throw the molecules out of the trap \cite{Tokunaga2011}. An ac trap has been developed that confines high-field seeking molecules by rotating a saddle-shaped potential \cite{VanVeldhoven2005}, as is done in an rf ion trap. However, these molecular traps are shallow and collisions transfer molecules from stable to unstable trajectories \cite{Tokunaga2011}. Optical dipole traps can also confine ground-state molecules, but they too are shallow and they have small volumes. More suitable is a microwave trap, as suggested in \cite{DeMille2004}, where ground-state molecules are confined near the maximum intensity of a microwave field. Using a microwave cavity with realistic Q-factor and input power, a trap depth of about 1\,K is feasible for a wide range of molecules. The cavity can have an open structure to provide good access to the molecules, the trap has a large volume, and the confined molecules are stable against collisions making the trap suitable for sympathetic or evaporative cooling.

\begin{figure}\label{fig:cavityDiagram}
  \centering
  \includegraphics[width=0.5\textwidth]{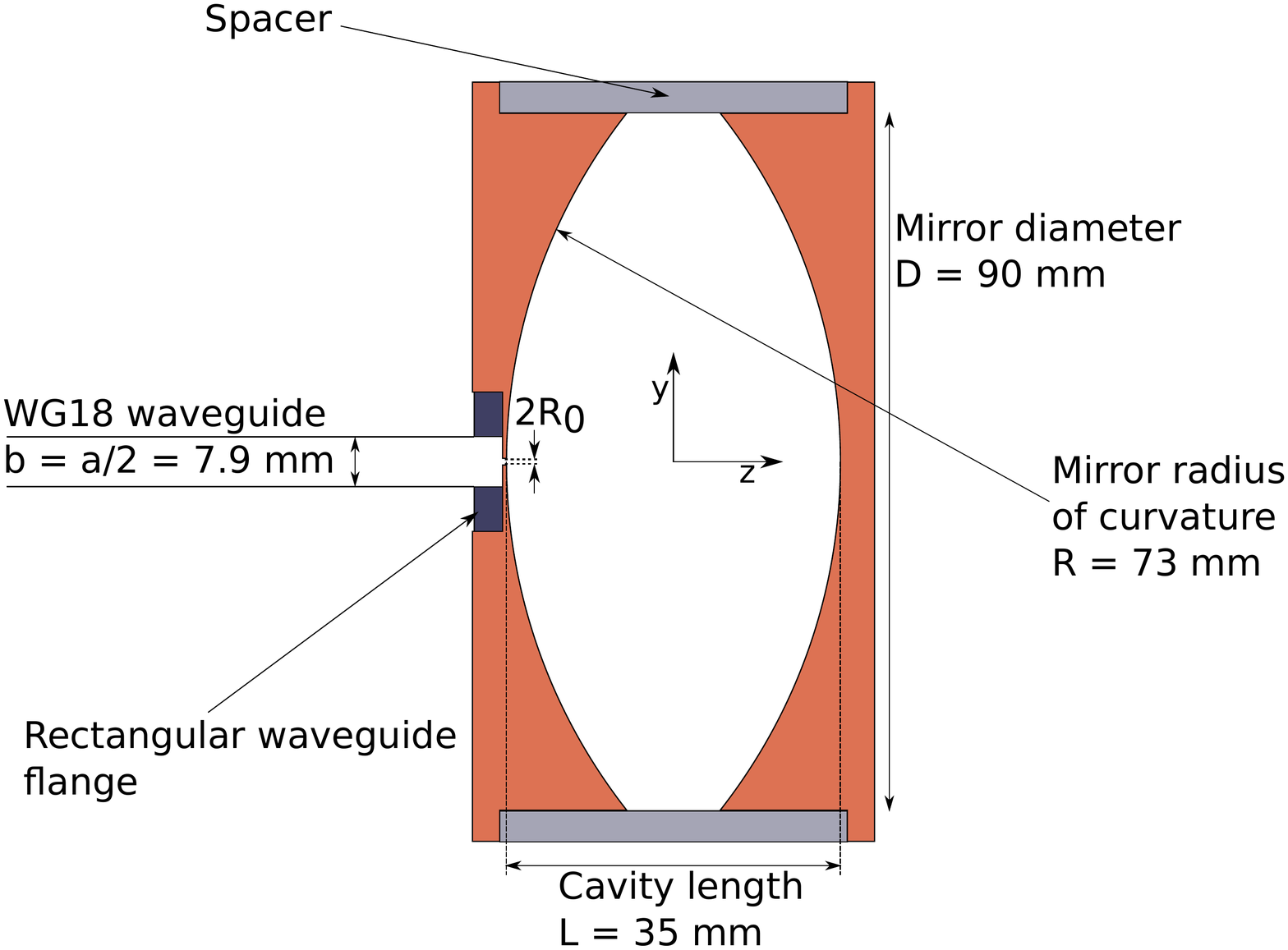}
  \caption{Cavity used in the experiment. There are holes in the spacer for optical access.}
\end{figure}

Here, we explore how to make a microwave trap using a Fabry-P\'erot resonator. The deepest trap requires a cavity Q-factor as high as possible, a cavity mode as small as possible, and high power coupled efficiently into the cavity. Figure \ref{fig:cavityDiagram} shows the cavity we use. It has a separation of $L=36$\,mm between two identical copper mirrors with radius of curvature $R_{m}=73$\,mm and diameter $D=90$\,mm. Holes in the spacer provide good access to the centre of the cavity. The cavity is fed from a waveguide via a small hole in one mirror, radius $r_{h}$ and thickness $t=0.7$\,mm. The design raises a number of questions. Is it possible to obtain efficient coupling into the cavity this way, while maintaining a high Q-factor? What is the optimum hole size? What are the smallest transverse modes supported by the cavity, how does this depend on the hole size, and how do these modes compare with the ideal TEM$_{00}$ modes? How high a Q can be reached and how does this depend on mirror temperature? While there is a great deal of literature on microwave resonators, we did not find any previous comprehensive study of these questions. We answer them using a combination of theory, measurement, and numerical simulation.

\section{Theory}\label{sec:theory}

We begin with a theoretical study of the cavity geometry of figure \ref{fig:cavityDiagram}. We use a coordinate system with origin at the centre of the cavity, oriented with the cavity axis along $z$, the waveguide's long dimension (length $a$) along $x$, and its short dimension (length $b$) along $y$.

\subsection{Modes}

In the limit of a very small hole and sufficiently large mirrors, the resonant modes of the cavity are the set of Hermite-Gaussian modes. Since we wish to minimize the spot size, we concentrate on the Gaussian modes with no transverse excitation whose intensity distribution is simply
\begin{equation}
I=\frac{I_{0}}{1+z^{2}/z_{0}^{2}}e^{-2(x^{2}+y^{2})/w^{2}},
\end{equation}
where $w=w_{0}\sqrt{1+z^{2}/z_{0}^{2}}$ is the spot size, $w_{0}=\sqrt{\lambda z_{0}/\pi}$ is the minimum spot size, and $\lambda$ is the wavelength. The wavefront has radius of curvature, $R=z+z_{0}^{2}/z$. For the mode to resonate, the wavefront at the mirror must match the mirror curvature, and this determines the Rayleigh parameter $z_{0}=L/2\sqrt{2R_{m}/L - 1}$. For the parameters of our cavity, the spot size at the cavity centre is $w_{0}=14.7$\,mm and the spot size at the mirrors is $w_{m}=17.0$\,mm. The frequencies of the modes are
\begin{equation}
\nu_{n} = \frac{c}{2L}\left(n+\frac{1}{\pi}\cos^{-1}(1-L/R_{m})\right).
\end{equation}
We focus on the mode with $n=3$ which has a resonant frequency near 14\,GHz.

\subsection{Quality factor}

The quality factor of the cavity is
\begin{equation}
Q=2\pi \nu_{n} \frac{2L}{c} \frac{1}{\beta},\label{eq:Q}
\end{equation}
where $\nu_{n}$ is the resonance frequency and $\beta$ is the fraction of power lost from the cavity in each round trip. At normal incidence, the fractional power loss due to reflection by a metal surface of resistivity $\rho$ is $\beta_{r} = 4\sqrt{\pi \nu_{n} \epsilon_0 \rho}$. When these resistive losses dominate over all other losses, the round-trip loss is $\beta = 2\beta_{r}$. A convenient expression for the resistivity is the Bloch-Gr\"{u}neisen formula,
\begin{equation}
\rho(T) = \rho_{0} + \frac{C}{m \Theta} \left(\frac{T}{\Theta}\right)^{5}\int_{0}^{\Theta/T}\frac{z^{5} e^{z}}{(e^{z}-1)^{2}}\,dz,
\label{eq:rho}
\end{equation}
where $m=63.5$ is the atomic mass of copper, $\Theta=343.5$\,K is the Debye temperature of copper, and $C=1.85\times 10^{-3}$\,$\Omega$\,m\,K$^{-1}$ is a normalization constant chosen so that $\rho(293\,K) = 1.68 \times 10^{-8}$\,$\Omega$\,m \cite{Matula1979}. The quantity $\rho_{0}$ is the residual resistivity of the metal at 0\,K and depends on the concentration of impurities and defects in the metal. It is common to describe a particular sample by its residual resistivity ratio, ${\rm RRR}=\rho(300\,K)/\rho_{0}$. For high purity copper, RRR values of several hundred are typical, in which case $\rho_{0}$ makes a negligible contribution to the resistivity over the entire temperature range of interest here. At room temperature, we expect $\beta_{r} = 3.2\times 10^{-4}$ and an associated Q-value of 32500.

The diffraction loss due to spillover of the mode around the edges of the mirrors is $e^{-D^{2}/2w_{m}^{2}}$, for a perfect Gaussian mode. For our parameters, this is only $7.6 \times 10^{-7}$, negligible compared with the resistive loss.

\subsection{Coupling via a small hole}

Next we consider the coupling from the waveguide into the cavity, via the hole. Our aim is to find the amplitude and phase of the field reflected from the cavity, and their dependence on frequency and hole size, to compare against our measurements.

Where the waveguide meets the cavity the field is partly reflected back into the waveguide and partly transmitted into the cavity. To find expressions for the reflected and transmitted fields, we draw on the previous analysis of Mongia \cite{Mongia1993}. The coupling hole is much smaller than the wavelength and can be modelled as an oscillating electric dipole perpendicular to the hole and an oscillating magnetic dipole in the plane of the hole, as shown by Bethe \cite{Bethe1944}. These dipoles are driven by the incident field. The waveguide propagates the TE$_{10}$ mode, which has no longitudinal component of electric field, so the electric dipole is not excited. The magnetic dipole is excited by the component of the incident magnetic field in the $x$-direction. In Bethe's theory, the effective magnetic moment of the hole is ${\bf m}_{h}=\alpha_{\mathrm{M}} {\bf H}$, where $\alpha_{\mathrm{M}}=\case{4}{3} f_t r_{h}^{3}$ is the magnetic polarizability of the hole and $\bf{H}$ is the magnetic field at the hole. The factor $f_{t}$ accounts for the thickness, $t$, of the hole and its value is calculated in \cite{McDonald1972, Levy1980} to be
\begin{equation}
f_t = \exp\left(-\frac{2\pi A_{m} t}{\lambda_c}\sqrt{1-\left(\lambda_c/\lambda\right)^{2}}\right),
\end{equation}

\noindent where $\lambda_c = 3.412 r_{h}$ and $A_{m}t=1.0064t + 0.0819r_{h}$ for all $t>0.2r_{h}$. For our $t=0.7$\,mm, this correction factor is 0.24 when $r_h=1$\,mm and 0.54 when $r_{h}=2.5$\,mm.

To find an accurate result for ${\bf m}_{h}$, the driving field $\bf{H}$ must include the fields produced in the waveguide and the cavity by the hole, as well as the incident field \cite{Cohn1952, Collin1991, Collin2001}, and so we write
\begin{equation}\label{eq:holeMagneticDipole}
  m_{h} = \alpha_{\mathrm{M}}\left(H_{i}+H_{w}-H_{c}\right).
\end{equation}

\noindent Here, $H_{i}$ is the incident magnetic field that would be present in the absence of the hole, $H_{w}$ is the field induced in the waveguide by the hole, and $H_{c}$ is the field induced in the cavity by the hole. All components are in the $x$ direction and are evaluated at the centre of the hole. In the absence of the hole, the incident field is a standing wave of the waveguide's $\mathrm{TE}_{10}$ mode due to reflection from the conducting wall at the end of the waveguide. We choose a coordinate system and normalization of the field launched into the waveguide such that $H_{i}=2$. Following section 4.13 of Collin \cite{Collin2001}, the field induced in the waveguide by the hole is found to be
\begin{equation}
  H_{w} = -\frac{2\rmi k_{g} m_{h}}{a b},\label{eq:wgField}
\end{equation}

\noindent where $k_{g}$ is the wavevector in the waveguide and $a$ and $b$ are the waveguide dimensions. Following section 7.9 of Collin \cite{Collin2001}, the field induced in the cavity by the hole is
\begin{equation}
  H_{c} = \frac{\nu^{2} m_{h} H_{n}}{\nu_{n}^{2}-\nu^{2}\left[1+\left(1-\rmi\right)/Q_{0}\right]} H_{n}, \label{eq:fieldExpansion}
\end{equation}

\noindent where $\nu$ is the microwave frequency, $\nu_{n}$ is the frequency of the resonant mode, $Q_{0}$ is the cavity quality factor for this mode (in the absence of the hole), and $H_{n}$ is the field of the resonant mode, normalized so that $\int H_{n}^{2} dV = 1$. To evaluate this volume integral, we approximate the wavefronts as plane waves, and thus obtain

\begin{equation}
H_{\mathrm{c}}=\frac{-\nu^{2}m_{h}}{\nu_{n}^{2}-\nu^{2}\left[1+\left(1-\rmi\right)/Q_{0}\right]}\frac{4}{\pi w_{\mathrm{m}}^{2}L}. \label{eq:cavityField}
\end{equation}

\noindent Using these results, we find the magnetic moment of the hole to be
\begin{equation}
  m_{h}=\frac{2\alpha_{\mathrm{M}}}{1+2\rmi c_{2} - \frac{c_{1}\nu^{2}}{\nu_{n}^{2}-\nu^{2}\left(1+(1-\rmi)/Q_{0}\right)}}.
  \label{eq:mh}
\end{equation}

\noindent where $c_{1}=4\alpha_{\mathrm{M}}/\left(\pi w_{m}^{2}L\right)$ and $c_{2}=\alpha_{\mathrm{M}}k_{g}/a b$ are dimensionless constants. The field reflected by the cavity is the sum of the field reflected in the absence of the hole and the field produced in the waveguide by the hole. Using equations (\ref{eq:wgField}) and (\ref{eq:mh}), this reflected field is

\begin{equation}
H_{r} = 1 + H_{w} = 1 - \frac{4 \rmi c_{2}}{1+2\rmi c_{2} - \frac{c_{1}\nu^{2}}{\nu_{n}^{2}-\nu^{2}\left(1+(1-\rmi)/Q_{0}\right)}}.
\label{eq:reflectedField}
\end{equation}

When characterizing the cavity experimentally, we measure the cavity reflection coefficient ${\cal R} = |H_{r}|^{2}$, and the phase $\theta = \textrm{arg}(H_{r})$. From the above expression, and after a great deal of excruciating algebra, we find the reflection coefficient to be

\begin{equation}
{\cal R} = |H_{r}|^{2} = 1 - \frac{4 \kappa \nu_{n}^{2}/(2Q_{0})^{2}}{(\nu-\nu_{n}+\delta)^{2}+(1+\kappa)^{2}\nu_{n}^{2}/(2Q_0)^{2}},
\label{eq:reflectionCoeffcient}
\end{equation}
where $\kappa=2Q_{0}c_{1}c_{2}/(1+4c_{2}^{2})$ and $\delta = \left(1+\frac{\kappa}{2c_{2}}\right)\frac{\nu_{n}}{2Q_{0}}$. In deriving this result, we have made the approximation $(\nu^{2}-\nu_{n}^{2}) \simeq 2\nu_{n}(\nu-\nu_{n})$, and used the fact that $Q_{0} \gg 1$. For all the hole sizes used in the present work, $c_{2}^{2} \ll 1$, and we take
\begin{equation}
\kappa \simeq 2Q_{0}c_{1}c_{2} \simeq \frac{8Q_{0} \alpha_{\mathrm{M}}^{2} k_{g}}{\pi w_{m}^{2} L a b}.
\end{equation}

We see from equation (\ref{eq:reflectionCoeffcient}) that the dependence of the reflected intensity on the microwave frequency follows a Lorentzian distribution. The resonance frequency is shifted to a lower frequency than $\nu_{n}$, by the amount $\delta$. The first part of this frequency shift depends only on the Q-factor and is the usual shift of an oscillator's resonance due to damping, while the second part depends on the size of the coupling hole and is due to the perturbation of the mode by the hole. At the reflection minimum where $\nu = \nu_{n}-\delta$, the fraction of incident power coupled into the cavity is $4\kappa/(1+\kappa)^{2}$, which is unity when $\kappa = 1$. This condition gives us the hole size for critical coupling as
\begin{equation}
r_{h}^{\kappa=1} = \left(\frac{9 \pi w_{m}^{2} L a^{2}}{256 Q_{0} k_{g} f_{t}^{2}}\right)^{1/6}.
\label{eq:criticalHole}
\end{equation}
\noindent We have included the thickness correction factor $f_{t}$ in this expression. It depends on $r_{h}$ but the dependence is quite weak and so the correct values of both $f_{t}$ and $r_{h}^{\kappa=1}$ are easily found in just a few iterations. For our cavity parameters, we predict $r_{h}^{\kappa=1}=2.35$\,mm. When the Q-factor is high, it is equal to the frequency of the resonance divided by its full width at half maximum. Using this relation, we see from equation (\ref{eq:reflectionCoeffcient}) that the Q-factor in the presence of the hole (known as the loaded Q-factor) is $Q_L = Q_{0}/(1+\kappa)$, which is equal to $Q_{0}/2$ when the cavity is critically coupled.

From the argument of $H_{r}$ in equation (\ref{eq:reflectedField}) we find the phase of the reflected field to be
\begin{equation}
\theta = \tan^{-1}\left(\frac{-4c_{2}\left[(\nu - \nu_{n} + \delta - \delta')^{2}-\delta'^{2}\right]}{(\nu - \nu_{n} + \delta)^{2}+(1-\kappa^{2})\nu_{n}^{2}/(2Q_0)^{2}}\right),
\label{eq:phase}
\end{equation}
\noindent where $\delta'= \kappa \nu_{n}/(8c_2 Q_{0})$. In deriving this expression, we have once again set $(\nu^{2}-\nu_{n}^{2}) \simeq 2\nu_{n}(\nu-\nu_{n})$ and have used the fact that $Q_{0} \gg 1$ and $c_{2}^{2} \ll 1$.

\section{Methods}

\subsection{Experiment}

The mirrors drawn in figure \ref{fig:cavityDiagram} were machined from 99.99\% pure oxygen-free copper. After machining and polishing, they were immersed in a 1\,M solution of ammonium persulfate for one minute to etch away oxides and impurities, and passivate the surface \cite{Rosebury1993}. A standard WG18 rectangular waveguide is bolted into an inset machined into the back of one mirror. This delivers the microwave power which is coupled into the cavity via the circular coupling aperture in the centre of the mirror. The coupling hole was machined using a wire eroder to avoid mechanical deformations of the mirror. The hole radius was varied from 1\,mm to 4\,mm, and was measured with 5\,$\mu$m precision. A coordinate measuring machine was used to measure the radius of curvature of the mirrors and the thickness of the coupling hole.

For each hole size, the intensity and phase of the field reflected from the cavity was measured as a function of frequency using a vector network analyzer (VNA, Anritsu 37247C) scanned between 13.85 and 14\,GHz. From this data the resonance frequency, transmission and quality factor were determined.

The field profile of the mode in the cavity was measured using a bead-pull technique similar to that used by Battaglia \etal \cite{Battaglia1970}. A dielectric bead, small compared with the microwave wavelength, is pulled through the cavity causing a small perturbation. The fractional shift in the resonance frequency is $\delta\nu/\nu_{n} = \zeta I/I_{0}$, where $I$ is the intensity at the position of the bead, $I_{0}$ is the intensity at the cavity centre, and $\zeta$ is a proportionality constant that depends on the geometry of the cavity and on the volume and dielectric constant of the bead, but not on its position. We used a PTFE bead of diameter 1.59\,mm, with a 0.4\,mm diameter hole drilled through, glued onto a 0.15\,mm diameter nylon thread. We locked the microwave source onto the cavity resonance and pulled the bead through the cavity while monitoring the control voltage of the feedback loop. This control voltage gives the frequency shift due to the bead, and hence $ I/I_{0}$.

To test how the quality factor depends on temperature, the mirrors were attached to cooling blocks and the apparatus was housed in a vacuum chamber pumped to a pressure below $10^{-2}$\,mbar. Liquid nitrogen circulating through the blocks cools the mirrors to 77\,K, and a thermocouple attached to the back of one mirror measures the temperature.

\subsection{Simulation}

We used a commercial finite-difference time-domain package (CST Microwave Studio) to solve Maxwell's equations for the cavity geometry shown in Figure \ref{fig:cavityDiagram}. In these simulations the spacer was omitted so that the cavity sides are completely open. The material properties of the waveguide and cavity were set to those of 99.99\% purity oxygen free copper at room temperature. The cavity was excited via a waveguide port placed at the end of a length of WG18 rectangular waveguide. The frequency-dependent amplitude and phase of the reflected field at this port were determined for a number of different coupling aperture sizes using the built-in frequency domain solver. The structure was placed on a tetrahedral mesh with $10^{-9}$ accuracy and adaptive tetrahedral mesh refinement. The cavity resonances were first found using a broad frequency sweep, and the lowest-order Gaussian mode with $n=3$ (TEM$_{003}$) was identified. Then, at this resonance frequency, the electric field magnitude was found everywhere inside the structure.

\section{Results}
\label{sec:results}

\subsection{Field profile}
\label{sec:fieldprofile}

\begin{figure}
  \centering
  \includegraphics[width=\textwidth]{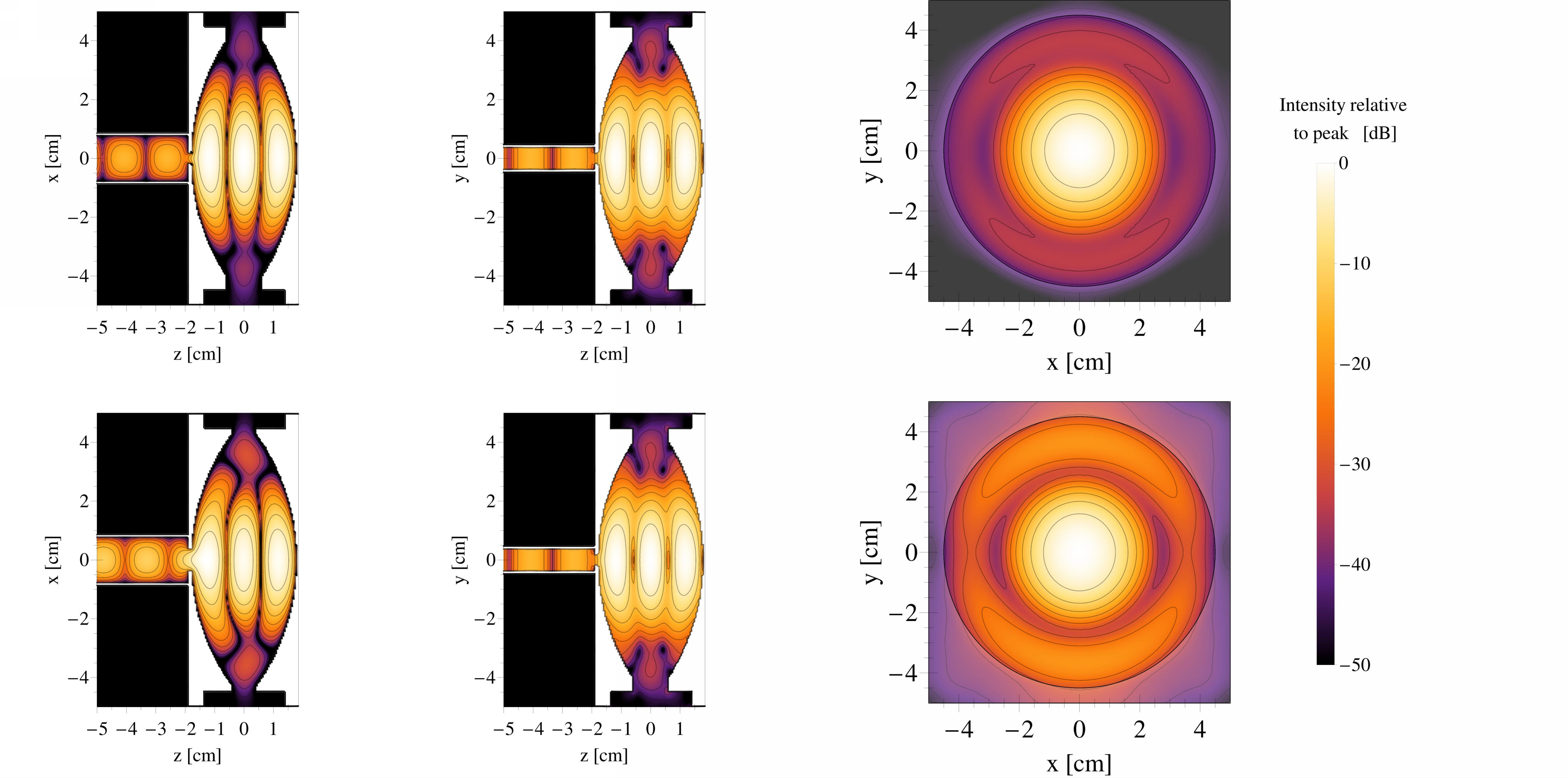}
  \caption{Cross sections of the simulated $\mathrm{TEM}_{003}$ mode in the xz, yz and xy-planes. Upper: $r_{h}=2.0$\,mm. Lower: $r_{h}=3.9$\,mm. The contours are spaced by -5dB.}
  \label{fig:longProfiles}
\end{figure}

Figure~\ref{fig:longProfiles} shows the simulated field profile in the xz, yz and xy-planes for the $\mathrm{TEM}_{003}$ mode and for two different hole sizes, $r_h=2.0$\,mm and 3.9\,mm. The profile in the xy-plane is particularly revealing. It shows that the mode resembles a Gaussian mode, but with an extra ring in the wings. For $r_h=2.0$\,mm the intensity in these wings is small, about -35\,dB relative to the peak intensity, but this increases as the hole size increases, reaching -20\,dB for $r_h=3.9$\,mm. To make a high-Q cavity we must minimize the spillover at the edges of the mirrors. For the ideal Gaussian mode in our cavity geometry, this diffraction loss is less than $10^{-6}$ - negligible compared to the absorption loss in the mirrors. The simulation shows that the actual power at the mirror edges is far larger, and we will see in section \ref{sec:Q} that this reduces the Q-factor below the expected value, especially for larger holes. We also see from figure~\ref{fig:longProfiles} that, while the cavity is axially symmetric, the mode is not. As shown in \cite{Svishchov2009}, the Fabry-P\'erot cavity has pairs of degenerate, axially asymmetric modes with orthogonal polarizations. The equal superposition of these modes has the expected axial symmetry. The rectangular waveguide excites only one of the degenerate modes, breaking the symmetry.

\begin{figure}[tb]
  \centering
  \includegraphics[width=0.8\textwidth]{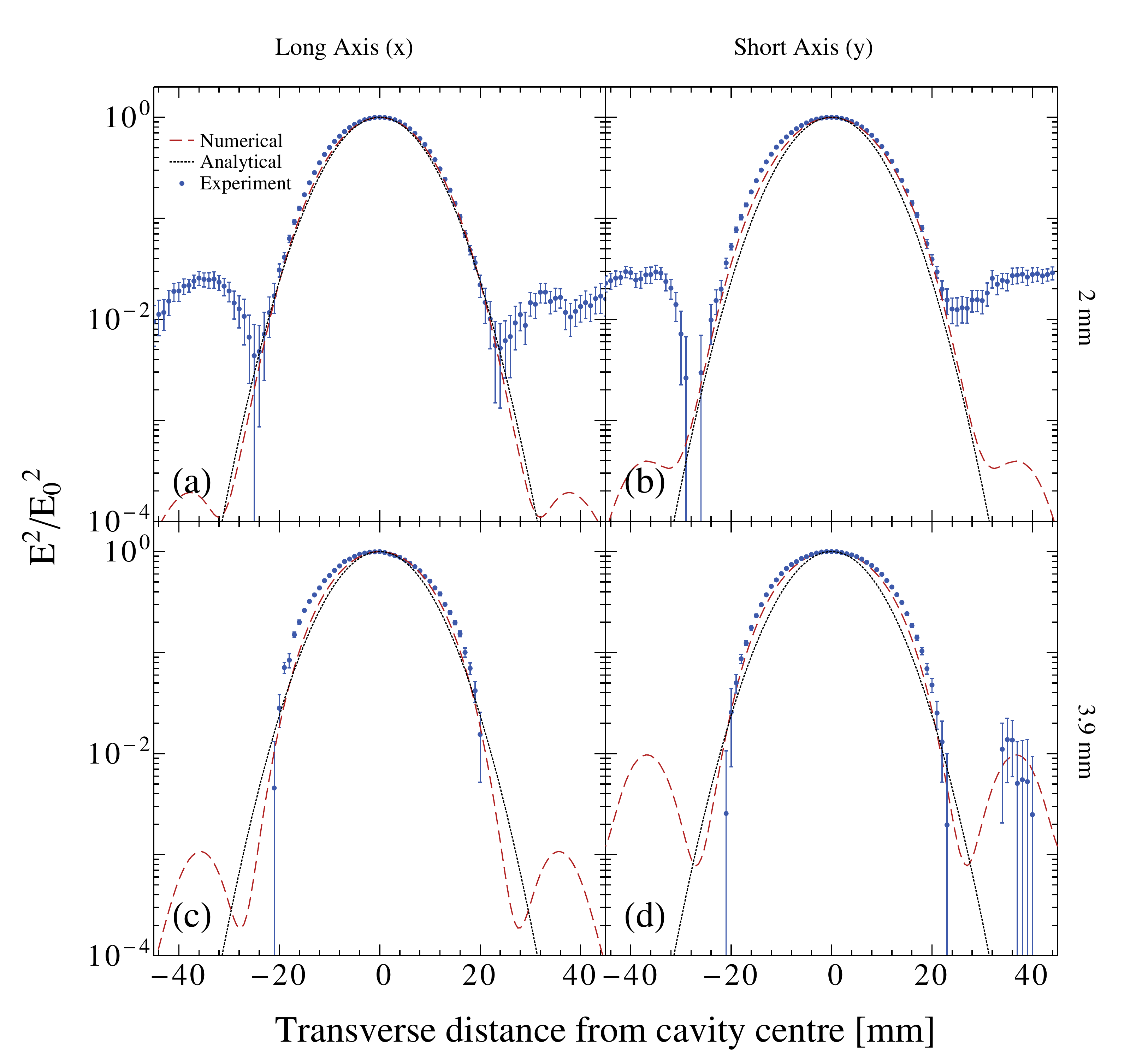}
  \caption{The transverse profile of the mode at the waist, measured using the bead-pull method and simulated with FDTD methods. The profiles parallel to both short and long dimensions of the waveguide are shown for coupling apertures with radius 2.0 mm and 3.9 mm.}
  \label{fig:beadPull}
\end{figure}

Figure~\ref{fig:beadPull} shows the intensity profiles along $x$ and $y$, measured at the centre of the cavity using the bead-pull method, and compares this to the simulated intensity profile and to the ideal Gaussian beam profile for this cavity. We first focus on the upper graphs where the hole size is $r_h=2.0$\,mm. In the $x$-direction, parallel to the waveguide's long dimension, the simulation predicts an intensity profile very similar to the ideal Gaussian profile all the way out to $y=30$\,mm, while in the $y$-direction the simulated profile is 10\% wider than the ideal Gaussian mode. At larger distances from the centre we see the extra ring of intensity discussed above. The experimental data follows closely the prediction of the simulation out to $r=20$\,mm, but beyond this distance the measured intensity is up to 100 times higher than predicted. This discrepancy is probably due to cross-coupling into another near-degenerate cavity mode, as discussed in more detail in section \ref{sec:Q}. Turning now to the larger hole size, $r_h=3.9$\,mm, the simulation predicts a slightly larger mode, particularly in the $y$-direction where it is 14\% wider than the ideal Gaussian mode. The measured intensity distribution is slightly wider again, 20\% wider than the ideal Gaussian. For this hole size we could only measure a few points in the wings of the distribution, but for these few points we find good agreement with the prediction of the simulation. The fact that the width of the cavity mode is only a little larger than in the ideal case is important for the goal of trapping atoms and molecules in the microwave field, since the electric field at the cavity centre is inversely proportional to this width.

\subsection{Quality factor and coupling efficiency}
\label{sec:Q}

\begin{figure}[tb]
  \centering
  \includegraphics[width=0.8\textwidth]{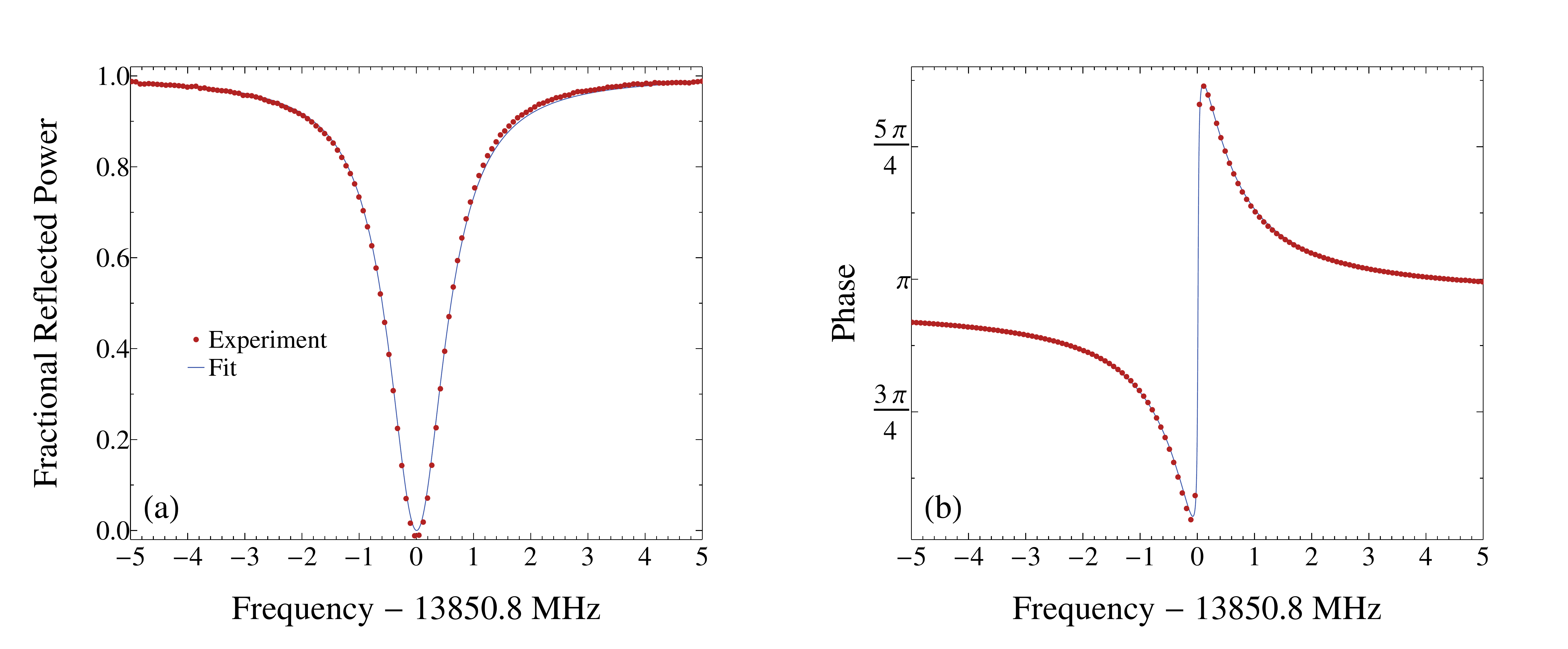}
  \caption{(a) Fractional reflected power and (b) relative phase as a function of microwave frequency, showing the TEM$_{003}$ resonance. Points: data. Lines: fits to equations (\ref{eq:reflectionCoeffcient}) and (\ref{eq:phase}).}
  \label{fig:resonance}
\end{figure}

Figure \ref{fig:resonance} shows how the power and phase of the field reflected from the cavity depend on the microwave frequency. Here the hole radius is 2.50\,mm and the temperature is 293\,K. We fit the reflected power and phase data to the models described by equations (\ref{eq:reflectionCoeffcient}) and (\ref{eq:phase}) respectively. The models fit the data well and from the fits we obtain an unloaded Q-factor of $Q_{0}=23150$ and a coupling coefficient of $\kappa = 0.959$. Note that $Q_0$ represents the quality factor that would be observed for $\kappa=0$, but since it is a function of diffractive losses it also depends on the hole size. 

\begin{figure}[tb]
  \centering
  \includegraphics[width=.8\textwidth]{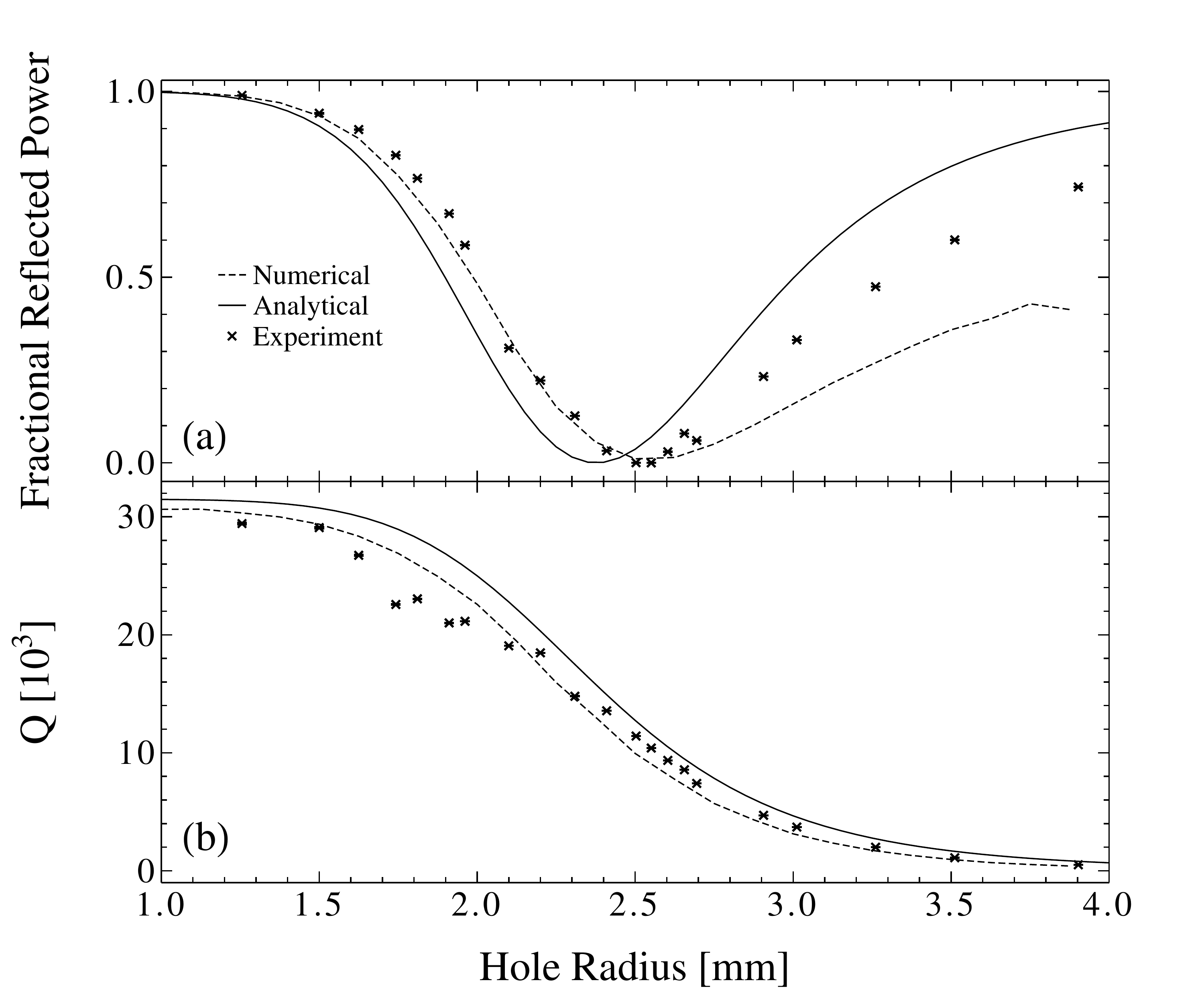}
  \caption{(a) Fractional reflected power and (b) quality factor as a function of the coupling aperture radius. Results are shown for the analytical model, numerical model, and actual cavity.}
  \label{fig:frpAndQ}
\end{figure}

Figure~\ref{fig:frpAndQ} shows how the measured reflection coefficient and the quality factor depend on the aperture radius. For very small holes, where the cavity is weakly coupled, the Q-factor is $Q_{0}=30000$ and almost all the power is reflected from the cavity. As the hole size increases the loaded Q-factor decreases and the reflected power reaches a minimum of less than 1\% at $r_h=2.55$\,mm where the cavity is critically coupled. As the hole size increases further the reflected power increases again while the Q-factor continues to decrease. At critical coupling, we might expect the measured Q-factor to be half of that which we measure with very small hole sizes, but we actually measure it to be a little over a third. This is because the electric field profile changes its shape as the hole size changes, bringing additional diffraction losses for larger hole sizes as discussed above.

Figure~\ref{fig:frpAndQ} also shows the predictions of the analytical theory described in section \ref{sec:theory}, and of the numerical simulations. Across the whole range of hole sizes, the numerical model predicts a smaller Q-factor than the analytical one. This difference is because there is more intensity in the wings of the mode than in the ideal Gaussian mode (see section \ref{sec:fieldprofile}), and so the diffraction loss around the edges of the mirrors is greater. The fractional discrepancy increases with increasing hole size because the intensity in the wings is larger when the hole is larger. The measured Q agrees well with the numerical simulation for most hole sizes, except when $r_h$ is between 1.6 and 2.0\,mm where the measured Q is up to 20\% smaller than calculated. We suggest an explanation for this below. The analytical theory of section \ref{sec:theory} predicts a lower reflected power than the numerical simulation when the cavity is under-coupled, and a higher reflected power when it is over-coupled. It also gives a smaller hole size for critical coupling. The reflected power measurements agree well with the numerical simulation up to critical coupling, and, importantly, they agree on the hole size required for critical coupling. For larger $r_h$, there are large differences between the measured reflected power and both calculations. Once again, these differences are due to the increased diffraction loss. To verify this, we repeated the FDTD simulation with a copper ring, inner diameter 86\,mm, added as a spacer to close off the cavity. With this addition, we find that the simulation result closely matches that of the analytical theory for both the Q-factor and the reflected power over the entire range of $r_h$. This shows that diffraction losses are responsible for the differences we observe when the cavity is open. In the experiment, we used a spacer between the two mirrors, with large gaps in the spacer to provide access into the cavity. This decreases the diffraction losses but does not eliminate them which explains why the measured reflected power of the over-coupled cavity lies half way between the analytical theory and the simulation of the open cavity.

\begin{figure}[tb]
  \centering
  \includegraphics[width=.8\textwidth]{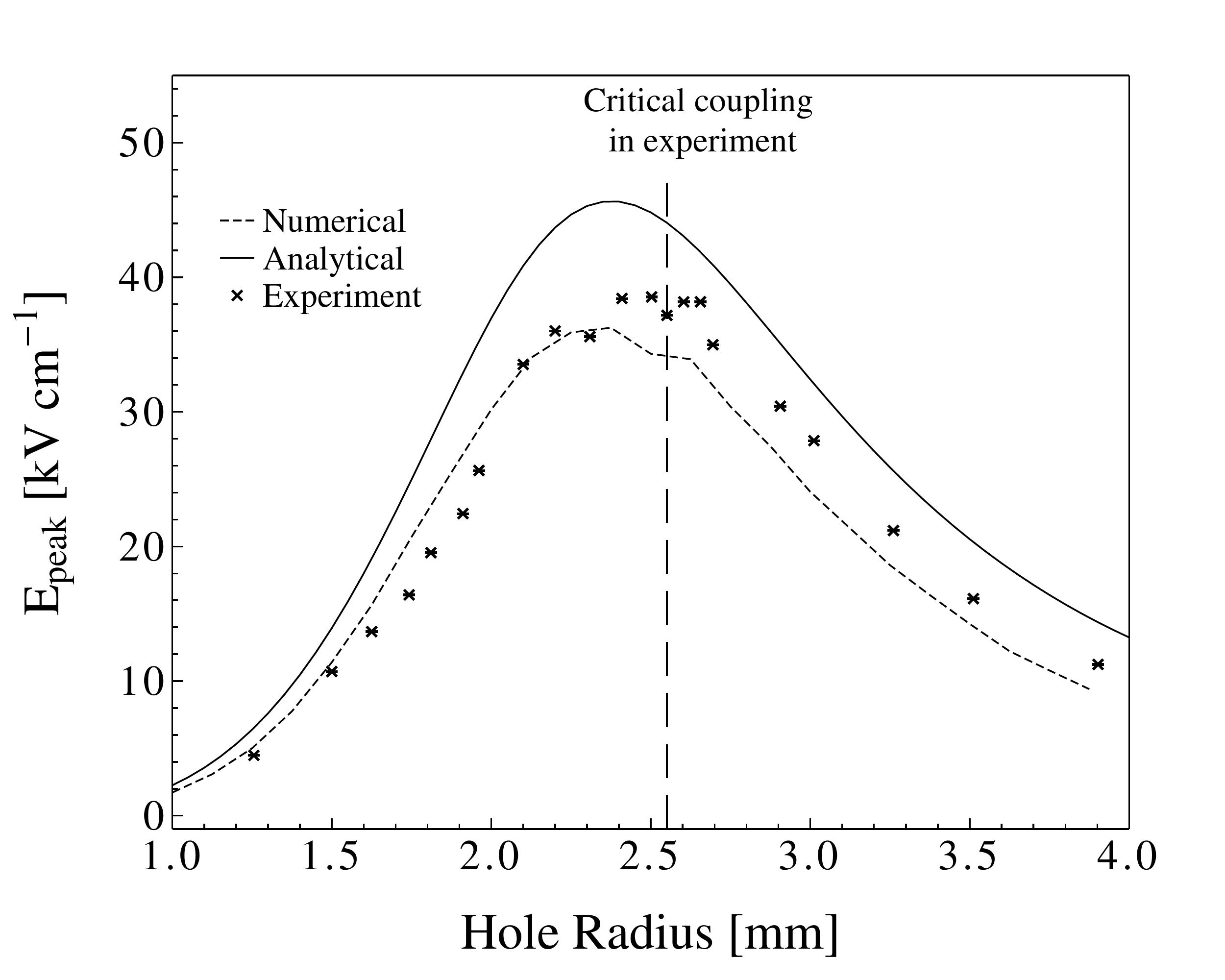}
  \caption{The peak electric field at the waist of the gaussian beam in the microwave cavity. Results are shown for the analytical model, numerical model, and actual cavity.}
  \label{fig:ePeak}
\end{figure}

The electric field at the centre of the cavity is
\begin{equation}
  E_{\mathrm{peak}}=\left(\frac{4 P_{0}(1-{\cal R}) Q_{0}}{\pi^{2}\nu\epsilon_{0}w_{0}^{2}L}\right)^{1/2},
\label{eq:ePeak}
\end{equation}
where $P_{0}$ is the forward power in the waveguide. For a microwave trap, we wish to maximize this electric field. Note that it is the unloaded quality factor, $Q_{0}$, that appears in equation (\ref{eq:ePeak}). In the ideal case, $Q_{0}$ does not depend on the hole size, but in reality its value decreases slightly as the hole size increases because of the increased diffraction loss. The quantity we measure is the loaded quality factor, $Q_{L}$, but we can determine $Q_{0}$ using $Q_{0}=Q_{L}(1+\kappa)$ with $\kappa$ found from the measurement of ${\cal R}$ and the relation $(1-{\cal R})=4\kappa/(1+\kappa)^{2}$ (see section \ref{sec:theory}). Figure~\ref{fig:ePeak} shows the electric field at the cavity centre determined from equation (\ref{eq:ePeak}) using the measured values for $Q_{0}$, ${\cal R}$ and $w_{0}$, and assuming $P_{0}=1.5$\,kW, a realistic value for a $\mathrm{K}_{\mathrm{U}}$ band klystron amplifier. The maximum field is obtained at critical coupling, where it reaches $E_{\mathrm{peak}}=39$\,kV/cm. Although the optimum hole size predicted by equation (\ref{eq:criticalHole}) gives a slightly under-coupled cavity, that choice of hole size (2.35\,mm) would still give a peak electric field of about 35\,kV/cm.

\begin{figure}[tb]
  \centering
  \includegraphics[width=.6\textwidth]{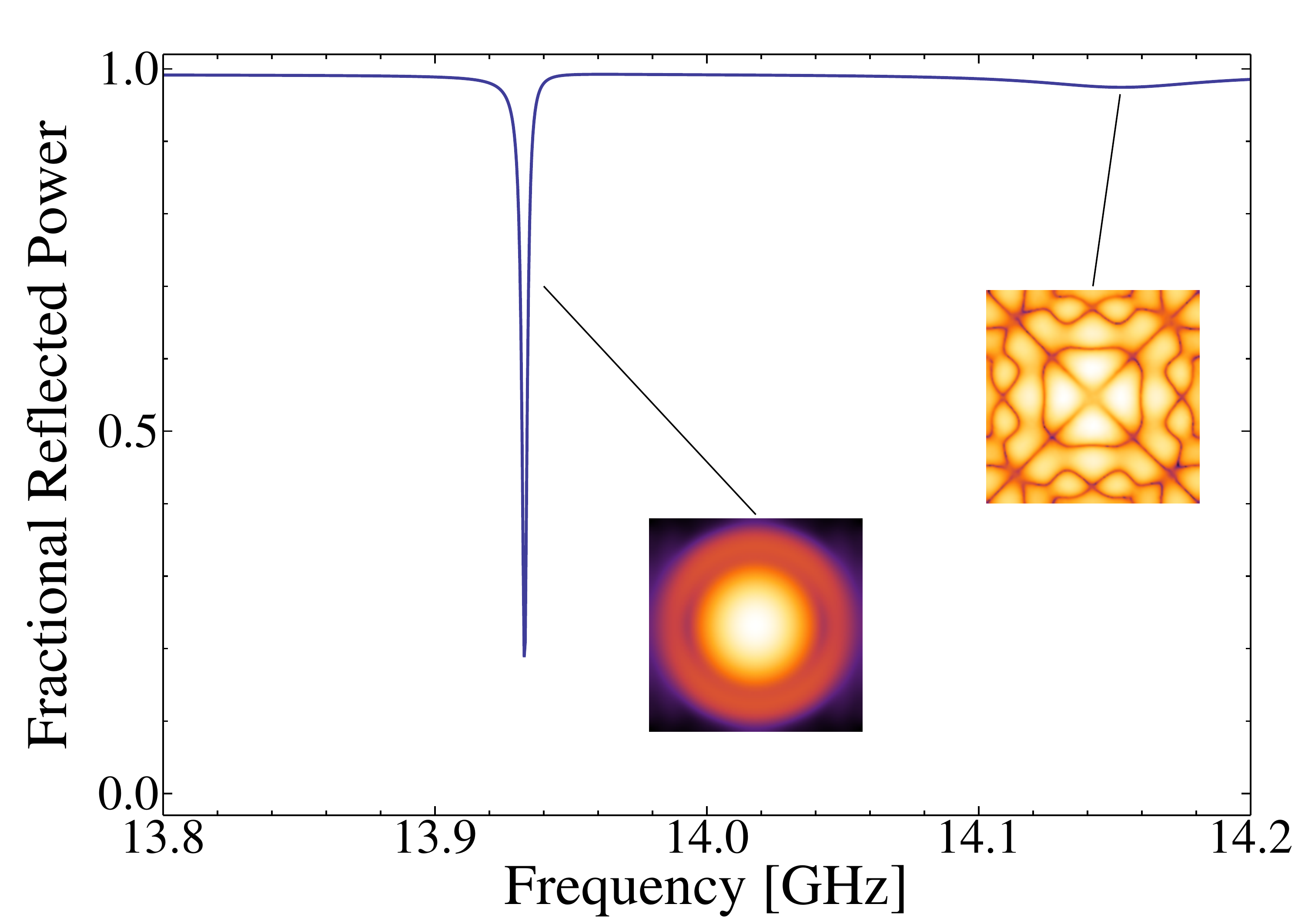}
  \caption{Simulated cavity spectrum near 14\,GHz for a hole size of $r_{h}=2.0$\,mm, showing the TEM$_{003}$ mode, and a higher-order mode whose Q-factor is much lower. The cavity length is slightly different to that used in the experiment, so that the modes can be clearly resolved. The insets show the intensity distributions of these two modes using the same logarithmic scale as in figure \ref{fig:longProfiles}}
  \label{fig:crossCoupling}
\end{figure}

We return now to the Q-factor of the cavity for hole sizes between 1.6 and 2.0\,mm, where there is a discrepancy between measurement and simulation (see figure~\ref{fig:frpAndQ}(b)). We suggest that this discrepancy is due to cross-coupling with a near-degenerate cavity mode of much lower Q. Figure \ref{fig:crossCoupling} shows the simulated spectrum of reflected power for a hole size of 2.0\,mm, showing the narrow TEM$_{003}$ resonance, and an additional higher-order mode which has a Q-factor of 185. The insets show the intensity distributions for these two modes. The higher-order mode has much of its intensity near the mirror edges, which is why the Q is low. We see this mode in the measured spectra, and have roughly confirmed this intensity distribution using the bead-pull method. As discussed in the context of equation (\ref{eq:reflectionCoeffcient}), the cavity resonances are shifted from $\nu_{n}$ by the amount
\begin{equation}
\delta = \left(\frac{1}{2Q_{0}} + \frac{2\alpha_{M}}{\pi w_{m}^{2}L}\right)\nu_{n}.
\end{equation}
This expression was found for the Gaussian mode, but the same expression will also apply to the higher-order modes except that $w_{m}$ will be replaced by a quantity characterizing the transverse extent of those modes. Recalling that $\alpha_{M}$ is proportional to $r_{h}^{3}$, and noting that the Gaussian mode has a much smaller $w_{m}$ than the higher-order mode, we see that the frequency of the Gaussian mode will shift far more rapidly with increasing hole size than the high-order mode. We see this in our simulations - the high-Q TEM$_{003}$ mode shifts by about 15\,MHz as the hole size increases from 1.0\,mm to 3.0\,mm, whereas the low-Q higher-order mode seen in figure \ref{fig:crossCoupling} shifts by less than 1\,MHz over this same range of hole sizes. As a result, the high-Q mode becomes degenerate with the low-Q mode for some particular hole size, which in the simulation is between 1 and 2\,mm. This corresponds approximately to the range of hole sizes where the measured Q-factor deviates from the expected value, so we attribute the lowered Q to cross-coupling between the two modes. This also explains the excess intensity measured in the wings of the profile for a 2.0\,mm hole size (see figure \ref{fig:beadPull}) - a small fraction of the power is coupled into the low-Q mode which has high intensity near the mirror edges. Mode coupling of this kind can be brought about by minor misalignments or imperfections in the cavity geometry and so can be missed by the numerical simulation.

\subsection{Temperature dependence of the quality factor}

We have seen that the cavity quality factor is close to the limit set by the surface resistivity of the mirrors. Cooling the mirrors should lower the resistivity, $\rho$, and so increase the quality factor.

\begin{figure}[tb]
  \centering
  \includegraphics[width=.6\textwidth]{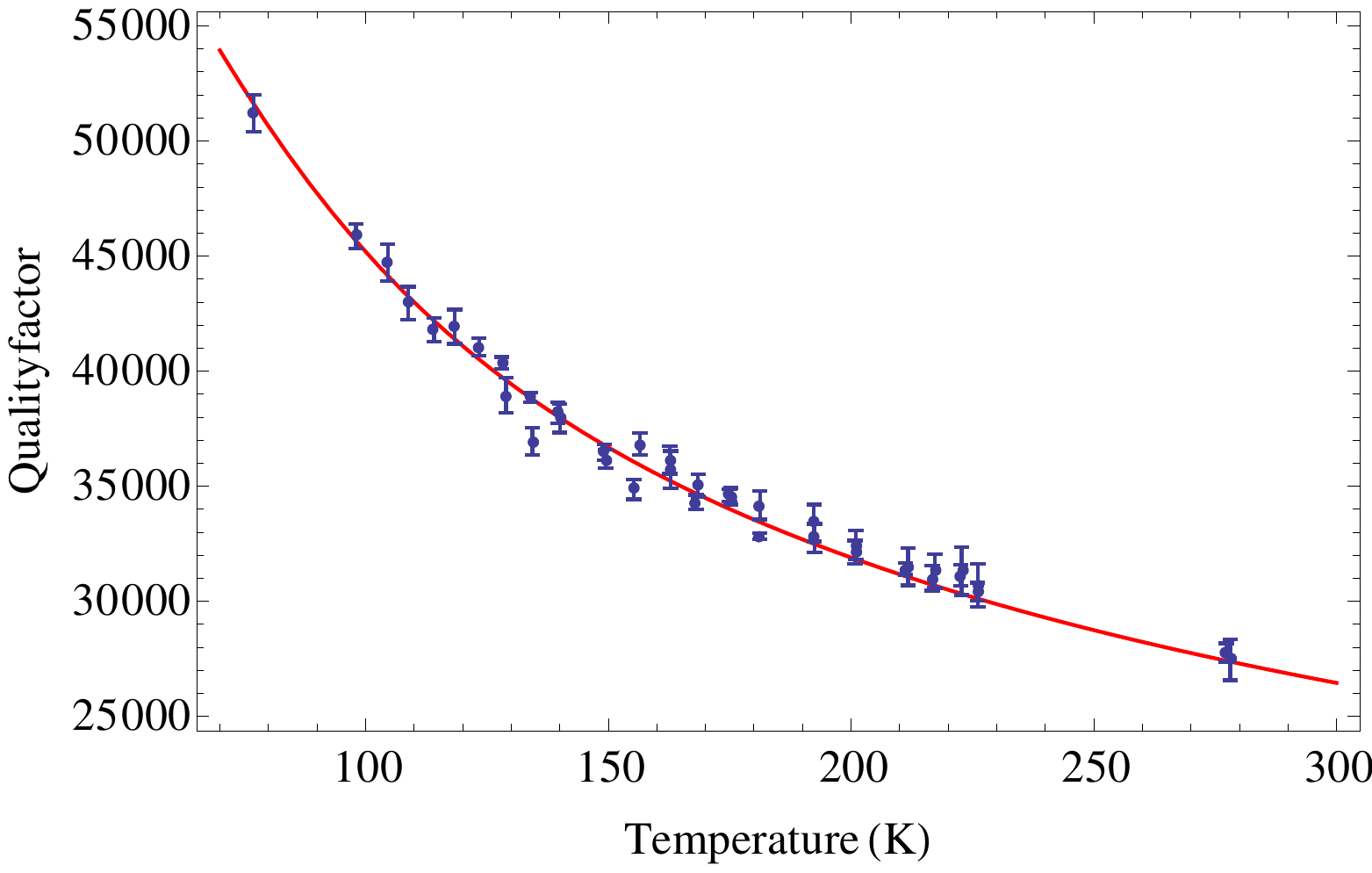}
  \caption{Quality factor of the cavity versus temperature. The solid line is a fit to the model discussed in the text.}
  \label{fig:temperature}
\end{figure}

Figure~\ref{fig:temperature} shows our measurements of the quality factor as a function of temperature. The Q-factor was measured as the cavity warmed up from 77\,K to room temperature. We used a mirror with a small hole to give a transmission of about 3\%, so that the hole contributes negligibly to the total round-trip loss and the intensity in the wings of the mode remains small.  We fit the data to equation (\ref{eq:Q}), with $\beta= 2\beta_{r} + \beta_{{\rm other}}$ and the temperature-dependent resistivity given by equation (\ref{eq:rho}). There are two free parameters in this fit, $\rho_{0}$ which is the residual resistivity and $\beta_{{\rm other}}$ which accounts for all non-resistive losses. The solid line in figure~\ref{fig:temperature} shows the best fit which gives $\rho_{0} = 2.1(3) \times 10^{-9}$\,$\Omega$\,m and $\beta_{{\rm other}}=8.8(7)\times 10^{-5}$.

It is common to describe a particular sample by its residual resistivity ratio, ${\rm RRR}=\rho(300\,K)/\rho_{0}$. For high purity copper, RRR values of several hundred are typical, whereas our fitted value of $\rho_{0}$ corresponds to ${\rm RRR}=8.1$. This is characteristic of a residual surface resistance that is much higher than that of the bulk material, which can be caused by a thin oxide layer on the surface, or by surface roughness. Using a surface profiler we measured a typical surface rms roughness of 1\,$\mu$m, about twice the skin depth at 14\,GHz. In a previous study \cite{Hernandez86}, this level of roughness was found to increase the room temperature resistivity by about 30\%, twice as much as we find here from our fit. We attribute $\beta_{{\rm other}}$ entirely to diffraction losses around the edges of the mirrors. We already noted in section \ref{sec:Q} that, even at room temperature and for a small hole, the measured Q is smaller than predicted by our simple model but is very close to the value found from the simulation. We found in the simulation that this extra loss vanishes when the cavity is closed, confirming that it is due to diffraction loss. We find that this diffraction loss, together with the temperature-dependent resistive loss and the residual surface resistance describes our data well over the whole temperature range.

\section{Discussion}

The depth of a microwave trap depends on the ac polarizability of the atom or molecule, and on the strength of the microwave electric field which we would like to make as high as possible. For example, a field of 30\,kV/cm at 14.5\,GHz gives a trap depth of 0.44\,mK for ground-state Li atoms and 400\,mK for ground-state CaF molecules. To produce such large fields requires a build-up cavity with high Q-factor, small mode size, and efficient coupling into the cavity. It is most convenient to couple microwave power directly from a waveguide into a cavity, and here we have shown that this can be done efficiently using a single coupling hole. In the presence of this hole, the intensity in the wings of the distribution increases well above the ideal Gaussian case. Nevertheless, the full width at half maximum of the intensity distribution at critical coupling is only 11\% wider than in the ideal case. The increased intensity in the wings increases diffraction losses and so reduces the Q-factor. In our simulations where the sides of the cavity are completely open, the Q-factor is reduced from the ideal case by about 30\% when the cavity is critically coupled. In our experiments, the sides were partly closed by a spacer between the mirrors, and our measured Q-factor at critical coupling is about 15\% less than the ideal case. The transmission of the hole increases with increasing radius and decreases with increasing thickness. To avoid excessive diffraction loss it is best to reach critical coupling with the smallest possible hole size, and so the hole should be kept as thin as possible. We have presented a formula for the required hole size for critical coupling, equation (\ref{eq:criticalHole}). In practice, we found the hole needed to be 8\% larger due to the extra diffraction loss. For certain hole sizes we observed a significant decrease of the Q-factor which we attribute to cross-coupling into a higher-order low-Q mode. Such cross-coupling can be avoided by a suitable choice of cavity length.

We were able to increase the quality factor of the cavity from 28000 to 51000 by cooling the mirrors to liquid nitrogen temperature. This increase was limited in part by surface resistance due to surface roughness, and in part by diffraction losses. For best performance, the rms surface roughness should be kept well below the skin depth, which for room temperature copper at 14.5\,GHz is 0.5\,$\mu$m.

Based on the present work, we have built a microwave trap for ground state atoms and molecules. It is designed to handle 1.5\,kW of input power at 14.5\,GHz while remaining at a temperature of 80\,K. In this case the electric field at the centre of the cavity will exceed 50\,kV/cm. For such large fields electric breakdown around the edges of the hole can be a problem. So far, we have coupled 700\,W into the cavity, limited by the cooling power, and did not observe any electrical breakdown.

\ack
We thank Jon Dyne, Steve Maine and Valerijus Gerulis for their expert technical assistance. This work was supported by the EPSRC and the Royal Society.


\section*{References}
\bibliographystyle{iopart-num}
\bibliography{references}

\end{document}